# Nonlinear dynamics of Airy-Vortex 3D wave packets: Emission of vortex light waves

Rodislav Driben[1, 2*] and Torsten Meier[1]

[1] Department of Physics & CeOPP, University of Paderborn, Warburger Str. 100, D-33098 Paderborn, Germany
[2] ITMO University, 49 Kronverskii Ave., St. Petersburg 197101, Russian Federation
*Corresponding author: driben@mail.uni-paderborn.de

The dynamics of 3D Airy-vortex wave packets is studied under the action of strong self-focusing Kerr nonlinearity. Emissions of nonlinear 3D waves out of the main wave packets with the topological charges were demonstrated. Due to the conservation of the total angular momentum, charges of the emitted waves are equal to those carried by the parental light structure. The rapid collapse imposes a severe limitation on the propagation of multidimensional waves in Kerr media. However, the structure of the Airy beam carrier allows the coupling of light from the leading, most intense peak into neighboring peaks and consequently strongly postpones the collapse. The dependence of the critical input amplitude for the appearance of a fast collapse on the beam width is studied for wave packets with zero and non-zero topological charges. Wave packets carrying angular momentum are found to be much more resistant to the rapid collapse, especially those having small width.

OCIS codes: Nonlinear optics, transverse effects in; (050.1940) Diffraction; (190.6135) Spatial solitons;(350.5500) Propagation; (190.3270)

Originally, Airy wave packets were obtained as solutions of the linear Schrödinger equation [1]. This finding inspired a significant interest in their counterparts in optical settings [2-5]. These fascinating self-accelerating light beams propagating along the bending trajectories were studied in the spatial and the temporal domains with multiple potential applications suggested [6-10]. Recently, observations of an electronic Airy beam [11] as well as Airy waves in plasmonic structures [12] have also been reported. Furthermore, multi-dimensional spatio-temporal Airy light bullets have also demonstrated [2, 13-17]. In the high power propagation regime the Airy wave structure gets distorted in the presence of a self-focusing Kerr nonlinearity, while the self-healing properties of Airy beams still leads to a strong resistance to the complete destruction [18].

In the temporal domain 1D temporal soliton [19-24] and multi-soliton [25] shedding out of Airy pulses were observed while the front of the Airy pulse continued its propagation along its bended trajectory. This dynamics occurs in the highly nonlinear regime via collisions of pulse peaks of which the Airy wave consists which leads to a subsequent transfer of energy and momentum from peak to peak until a soliton is released. This phenomenon can be viewed as an optical analog of Newton's cradle and it was demonstrated [26] to exist in other asymmetric light structures propagation regimes such as the evolution of dense multiple pulses with the same frequency and more importantly in fission of N-soliton under the dominant action of a third order dispersion.

Dynamics of vortices imposed into Airy wave structure have been studied in two dimensional settings [27-30] and it was demonstrated that the threshold of the rapid collapse [31-34] of 2D Airy waves carrying vortices is higher than that of the 2D Airy waves having no topological charge [30].

3D investigations would, however, be highly desirable to test predictions obtained in lower dimensions and, furthermore, the higher dimension adds additional degrees of freedom which may lead to new dynamical phenomena. The main objective of this letter is to demonstrate the emission of 3D waves in the course of the propagation of 3D Airy-Gauss light bullets in a nonlinear medium. We particularly show that if the initial light 3D light wave packet is launched with an initial topological charge (vorticity) the shed waves exhibit this vorticity too. In Kerr-type nonlinear media rapid collapse becomes a significant issue unlike in cases of dynamics with one-dimensional transverse coordinate. The mechanism can take place with input intensities below those leading to that rapid collapse. It will be shown that in a strongly nonlinear regime the structure of the Airy beam carrier allows to couple light from the most intense peak to its neighbors which prevents the usual rapid collapse. Furthermore, the introduction of a topological charge also strongly assists to postpone the collapse.

The propagation of the 3D light wave packets amplitude $u$ along the direction $z$ in nonlinear media obeys the 3D Nonlinear Schrodinger equation which reads in normalized form:

$$iu_z + (u_{tt} + u_{xx} + u_{YY}) + |u|^2 u = 0 \qquad (1)$$

Here, the diffraction in the x-y plane and the temporal dispersion act on the wave packet with effectively similar strength. In the low intensity regime with negligible nonlinearity the Airy-Bessel solution describes the dynamics of a linear light bullet [2, 13, 14]. Since we are interested in the dynamics in a strongly nonlinear case we study the evolution of a finite-energy light bullet with an input constructed by truncated temporal Airy and Gaussian functions in the x-y plane multiplied by components representing the toroidal vortex light distribution.

$$u_0 = A * \mathrm{Airy}(t)\exp(at)\exp(-\frac{x^2}{\omega^2})\exp(-\frac{y^2}{\omega^2})(x+iy)^S$$
(2)

Here, $A$ denotes the amplitude of the wave packet, w is the initial width in x- and y-direction, $a$ is the truncation of the Airy function, and S is the topological charge.

The evolution of the wave packet along z with input (2) governed by (1) is simulated using the Fourier transform split-step method in a 3D (x-y-t) domain of size $(8\pi)^3$, covered by a mesh of $256^3$ points. Fig. 1 demonstrates the evolution of the Airy-vortex wave packet with vorticity S=1. Three-dimensional snapshots (a-c) illustrate the shape of the wave packet intensity ($|u|^2$) at the level of $|u|^2 = 0.25$. The initial Airy-vortex structure seen in Fig.1 (a) gets distorted in the course of the propagation, see Fig.1 (b), in the strongly nonlinear regime with $A = 4.2$ and w = 2. The dynamics leads to light transfer from the leading peak to its neighbors via inter-pulse collisions, see Fig. 1(d), in analogy to a 1D temporal Airy pulse [19-24].

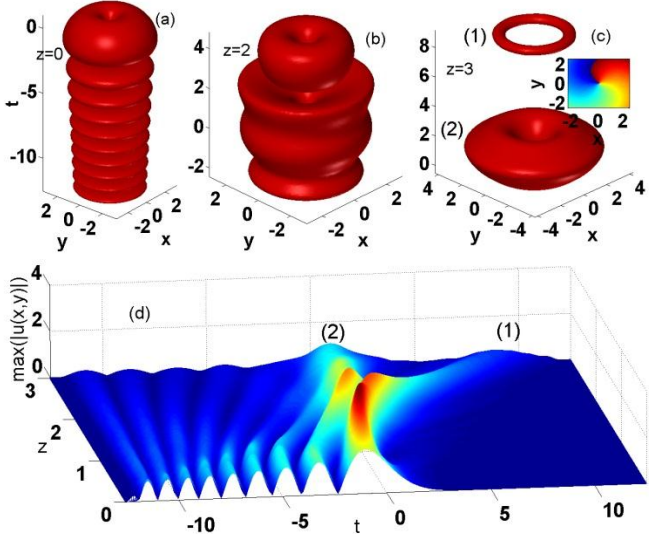

Fig. 1 (Color online) Snapshots describing the evolution of a wave packet taken (a) at the input (z = 0), (b) at the initial stage of the formation of the vortex to be emitted from the structure (z = 1), (c) after the emission of a strong vortex (z = 2). The inset in (c) displays the phase distribution at the maximum intensity of the emitted vortex. All snapshots are drawn at the intensity ISO level of $|u|^2 = 0.25$. (d) Evolution of the maximum of the wave packet - max($|u(x, y)|$) taken at the x - y plane. The input pulse amplitude is $A = 4.2$ and the width parameter-is w = 2. Leading Airy peak and the emitted light wave are marked by (1) and (2), respectively, in panels (c) and (d).

The absolute value of the maximum taken in the x-y plane is shown instead of the conventional intensity -$|u|^2$ in order to provide a better visibility of the dynamics in low intensity regions of the wave packet. The shed vortex that starts its formation at about z =1 becomes more powerful due to a course of collisions with Airy peaks until it is released from the structure. Figure 1(c), representing the wave packet at z = 3, demonstrates that at the intensity level of $|u|^2 = 0.25$ the decaying front of the 3D Airy wave packet located at t ~ 8.2 is rather weak, while more powerful emitted vortex centered at t ~ 1 appears to be much more pronounced. The phase distribution of the emitted light, measured at its center is shown by the inset in Fig. 1(c). It clearly corresponds to a vortex with topological charge S=1. We have marked the leading Airy peak by (1) and the emitted light by (2) both in Fig. 1(c) and Fig. 1(d) to clarify the the identification.

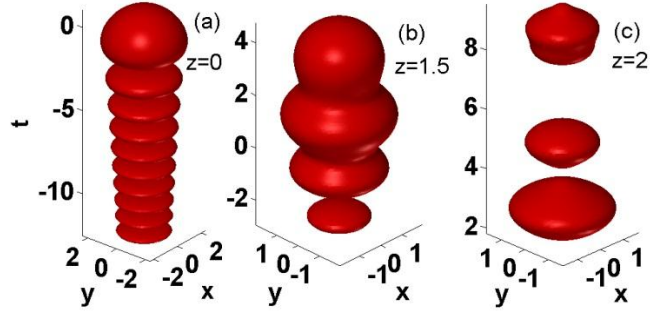

Fig. 2. (Color online) Snapshots describing the evolution of a wave packet with zero vorticity (S=0) taken (a) at the input (z = 0), (b) at the position just after the formation (z = 1.5) of the light waves emitted from the structure, and (c) after the emission of a strong light wave (z = 2). All the snapshots are drawn at the intensity ISO level of $|u|^2 = 0.15$. The input pulse amplitude is $A = 2.91$ and the width parameter-is w = 2.

The dynamics of the wave packets without topological charge (S = 0) and with charge S = 2 is shown in Fig. 2 and Fig. 3, respectively. These figures reveal a qualitatively similar nonlinear dynamics. Clearly, the emitted objects have the same topological charges as the injected wave packets.

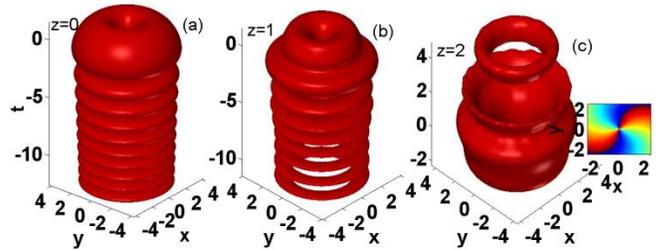

Fig. 3. (Color online) Snapshots describing the evolution of a wave packet with S=2, taken (a) at the input (z = 0), (b) at the position just before the formation (z = 1) of the vortex emitted from the structure, and (c) after the emission of a strong vortex (z = 2). All the snapshots are drawn at the intensity ISO level of $|u|^2 = 0.25$. The inputpulse amplitude is $A = 4.8$ and the width parameter-is w = 2.

In the case of still higher injected peak power the mechanism of light coupling from the most intense peak into neighboring peaks is not able to overcome the rapid collapse anymore. We can identify a critical peak power of the input wave packet for the appearance of the rapid collapse such as that shown in Fig. 4. The wave packet with the same width parameter w = 2, but with the initial amplitude $A = 4.2$ experiences a fast peak power growth after only a short propagation distance of z = 0.08. We have demonstrated the case of a wave packet with the vorticity S = 1, thus the collapse occurs at several points located on the circle perimeter of the field distribution in the x-y plane, as demonstrated in the inset of the Fig. 4.

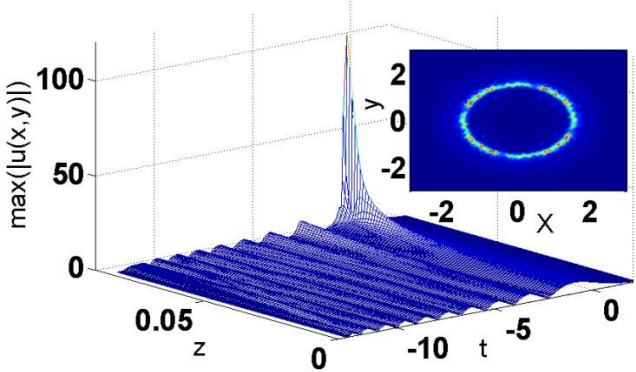

Fig. 4. (Color online) Typical example of the rapid collapse of the light packet with S=1 having an initial am amplitude A = 4.5 and width w = 2. Evolution in z-t plane is shown together with a x-y cross-section of the collapsing wave packet taken at z = 0.085.

The appearance of the rapid collapse can be controlled by the initial parameters of the wave packet, in particular, by its topological charge. Figure 5 (a) demonstrates the dependence of the critical input peak power $|u_0|^2_{cr}$ calculated from (2) on the square of the wave packet's width parameter - $w^2$. Values of critical power were obtained by running a set of direct simulations with increasing the input amplitude until the appearance of a clear rapid collapse such as shown in Fig. 4 rather than coupling of power into neighbor lobes such as in Fig. 1(d). One can clearly observe the postponing of the collapse due to the "charging" of the initial wave packet. Figure 5(b) also demonstrates such a comparison between the wavepackets with the angular momentum and those without the angular momentum in terms of total energy of the wavepacket.

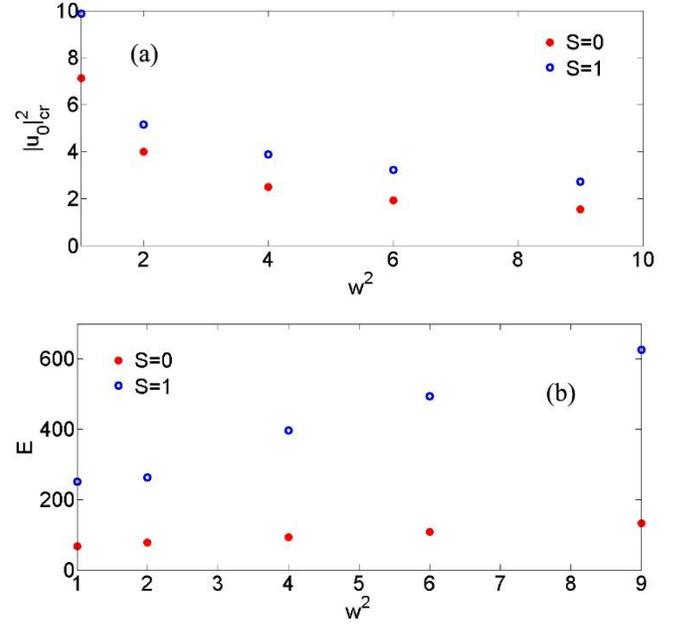

Fig. 5. (Color online) Critical input amplitude for the appearance of the rapid collapse of the wave packet as a function of the width squared ($w^2$). Red points represent the uncharged (S=0) case, while blue rings stand for wave packets with vorticity S=1.

Taking for example a wavepacket with the width parameter w = 2, the critical input peak power for S = 0 is $|u_0|^2_{cr}$ = 2.5, with the total energy $E_{cr}$ = 38. For S =1 the critical peak power is =93.5, with the total energy $P_{cr}$ =397.

In conclusion, we have studied the spatio-temporal evolution of 3D Airy-vortex wave packets in the strongly nonlinear regime under a threat of a rapid collapse. The structure of the Airy beam allows coupling of light from the leading, most intense peak into neighboring peaks and consequently strongly postpones the collapse. Emission of nonlinear 3D waves out of the main wave packet is demonstrated. The topological charges of the emitted objects are shown to be similar to those carried by the parental light structures. Wave packets carrying angular momentum are found to be much stronger resistant to immediate collapse than their uncharged counterparts from the point of view of the critical amplitudes.

In future works it will be interesting to consider nonlinear interaction of several 3D Airy-vortex wave packets, including interactions between wave packets carrying different topological charges.

Rodislav Driben wishes to express his gratitude to Boris A. Malomed for fruitful discussions. The authors gratefully acknowledge support provided by the Deutsche Forschungsgemeinschaft (DFG) via the Research Training Group (GRK 1464) and computing time provided by the PC² (Paderborn Center for Parallel Computing). Rodislav Driben gratefully acknowledges support provided the Government of the Russian Federation (Grant 074-U01) through ITMO Early Career Fellowship scheme.

References
[1] M. V. Berry and N. L. Balazs, Am. J. Phys. **47**, 264 (1979).
[2] G. A. Siviloglou and D. N. Christodoulides, Opt. Lett. **32**, 979 (2007).
[3] G. A. Siviloglou, J. Broky, A. Dogariu, and D. N. Christodoulides, Phys. Rev. Lett. **99**, 213901 (2007).
[4] T. Ellenbogen, N. Voloch-Bloch, A. Ganany-Padowicz, and A. Arie, Nature Photonics **3**, 395 (2009).
[5] I. Dolev, I. Kaminer, A. Shapira, M. Segev, and A. Arie, Phys. Rev. Lett. **107,** 116802 (2011).
[6] J. Baumgartl, M. Mazilu, and K. Dholakia, Nature Photonics **2, **675 (2008).
[7] P. Polynkin, M. Kolesik, J. V. Moloney, G. A. Siviloglou, and D. N. Christodoulides, Science **324**, 229 (2009).
[8] I. Kaminer, Y. Lumer, M. Segev, and D. N. Christodoulides, Opt. Express 19, 23132, (2011).
[9] Y. Hu, Z. Sun, D. Bongiovanni, D. Song, C. Lou, J. Xu, Z. Chen, and R. Morandotti, Opt. Lett., **37**, 3201 (2012).
[10] I. Chremmos, Opt. Lett., **39**, 15, 4611,(2014).
[11] N. Voloch-Bloch, Y. Lereah, Y. Lilach, A. Gover, and A. Arie, Nature **494**, 331 (2013).
[12] A. Minovich, A. E. Klein, N. Janunts, T. Pertsch, D. N. Neshev, Y. S. Kivshar, Phys. Rev. Lett. **107,** 116802 (2011).
[13] A. Chong, W. H. Renninger, D. N. Christodoulides, and F. W. Wise, Nature Photonics **4,** 103 (2010).
[14] P. Piksarv, H. Valtna-Lukner, A. Valdmann, M. Lõhmus, R. Matt, and P. Saari, Opt. Exp., **20**, 17220 (2012)
[15] D. Abdollahpour, S. Suntsov, D. G. Papazoglou, and S. Tzortzakis, Phys. Rev. Lett. **105**, 253901 (2010).
[16] W. P. Zhong, M. Belić, T. Huang, Phys. Rev. A **88**, 033824 (2013)
[17] P Piksarv, A Valdmann, H Valtna-Lukner, P Saari, Journal of Physics **497**,1 (2014)
[18] I. Kaminer, M. Segev, and D. N. Christodoulides, Phys. Rev. Lett.**106**, 213903 (2011)
[19] C.Ament, P. Polynkin, and J. V. Moloney, Phys. Rev. Lett. **107**, 243901 (2011).
[20] Y. Fattal, A. Rudnick, and D. M. Marom, Optics Express. **19**, 17298 (2011).
[21] R. Bekenstein and M. Segev, Optics Express 19, 23706, (2011).
[22] Y. Zhang, M. Belić, H. Zheng, H. Chen, C. Li, Y. Li, Y. Zhang, Optics express **22**, 7160, (2014)
[23] R. Driben, Y. Hu, Z. Chen, B.A. Malomed, and R. Morandotti, Optics Letters **38**, 2499 (2013).
[24] R. Driben, and T. Meier , Physical Review A **89**, 043817 (2014)
[25] Y. Zhang, M. Belić, Z. Wu, H. Zheng, K. Lu, Y. Li, Y. Zhang, Optics Letters **38**, 4585 (2013)
[26] R. Driben, B. A. Malomed, A. V. Yulin, and D. V. Skryabin, Phys. Rev. A **87**, 063808 (2013)
[27] M. Mazilu, J. Baumgartl, T. Cizmar, and K. Dholakia, Proc. SPIE 7430, 74300C–1 (2009)
[28] H. T. Dai, Y. J. Liu, D. Luo, and X. W. Sun, Optics Letters **35**, 4075 (2010).
[29] Y. Jiang, K. Huang, and X. Lu, Optics Express **20**, 18579 (2012).
[30] R. P. Chen, KH. Chew, and S.He, Scientific Reports **3**, 1406, (2013).
[31] Y. Silberberg, Optics Letters **15**, 1282 (1990).
[32] Malomed, B. A., Mihalache, D., Wise, F. and L. Torner, J. Opt. B **7**, R53 (2005).
[33] D. Mihalache, Rom. J. Phys. 57, 352 (2012).
[34] D. Mihalache, Rom. J. Phys. 59, 295 (2014).

Full References
[1] M. V. Berry and N. L. Balazs, "Nonspreading wave packets," Am. J. Phys. **47**, 264 (1979).
[2] G. A. Siviloglou and D. N. Christodoulides, "Accelerating finite energy Airy beams," Opt. Lett. **32**, 979 (2007).
[3] G. A. Siviloglou, J. Broky, A. Dogariu, and D. N. Christodoulides, "Observation of accelerating Airy beams," Phys. Rev. Lett. **99**, 213901 (2007).
[4] T. Ellenbogen, N. Voloch-Bloch, A. Ganany-Padowicz, and A. Arie "Nonlinear generation and manipulation of Airy beams," Nature Photonics **3**, 395 (2009).
[5] I. Dolev, I. Kaminer, A. Shapira, M. Segev, and A. Arie, "Experimental Observation of Self-Accelerating Beams in Quadratic Nonlinear Media," Phys. Rev. Lett. **107,** 116802 (2011).
[6] J. Baumgartl, M. Mazilu, and K. Dholakia, "Optically mediated particle clearing using Airy wavepackets," Nature Photonics **2,** 675 (2008).
[7]P. Polynkin, M. Kolesik, J. V. Moloney, G. A. Siviloglou, and D. N. Christodoulides, "Curved plasma channel generation using ultraintense Airy beams," Science **324**, 229 (2009).
[8] I. Kaminer, Y. Lumer, M. Segev, and D. N. Christodoulides, "Causality effects on accelerating light pulses," Opt. Express 19, 23132, (2011).
[9] Y. Hu, Z. Sun, D. Bongiovanni, D. Song, C. Lou, J. Xu, Z. Chen, and R. Morandotti, "Reshaping the trajectory and spectrum of nonlinear Airy beams," Opt. Lett. **37**, 3201 (2012).
[10] I. Chremmos, "Temporal cloaking with accelerating wave packets," Optics Letters, **39**, 15, 4611, (2014).
[11] N. Voloch-Bloch, Y. Lereah, Y. Lilach, A. Gover, and A. Arie, "Generation of electron Airy beams," Nature **494**, 331 (2013).
[12] A. Minovich, A. E. Klein, N. Janunts, T. Pertsch, D. N. Neshev, Y. S. Kivshar, "Generation and near-field imaging of Airy surface plasmons," Phys. Rev. Lett. **107,** 116802 (2011).
[13] A. Chong, W. H. Renninger, D. N. Christodoulides, and F. W. Wise, "Airy-Bessel wave packets as versatile linear light bullets," Nature Photonics **4,** 103 (2010).
[14] P. Piksarv, H. Valtna-Lukner, A. Valdmann, M. Lõhmus, R. Matt, and P. Saari "Temporal focusing of ultrashort pulsed Bessel beams into Airy–Bessel light bullets," Opt. Exp., **20**, 17220 (2012)
[15] D. Abdollahpour, S. Suntsov, D. G. Papazoglou, and S. Tzortzakis, "Spatiotemporal airy light bullets in the linear and nonlinear regimes," Phys. Rev. Lett. **105**, 253901 (2010).
[16] W. P. Zhong, M. Belić, T. Huang "Three-dimensional finite-energy Airy self-accelerating parabolic-cylinder light bullets" Phys. Rev. A **88**, 033824 (2013)
[17]P Piksarv, A Valdmann, H Valtna-Lukner, P Saari "Ultrabroadband Airy light bullets" Journal of Physics **497**,1 (2014)
[18] I. Kaminer, M. Segev, and D. N. Christodoulides, "Self-accelerating self-trapped optical beams," Phys. Rev. Lett.**106**, 213903 (2011)
[19] C.Ament, P. Polynkin, and J. V. Moloney, "Supercontinuum Generation with femtosecond self-healing Airy pulses," Phys. Rev. Lett. **107**, 243901 (2011).
[20] Y. Fattal, A. Rudnick, and D. M. Marom, "Soliton shedding from Airy pulses in Kerr media", Opt. Exp. **19**, 17298 (2011).
[21] R. Bekenstein and M. Segev, "Self-accelerating optical beams in highly nonlocal nonlinear media," Opt. Express 19, 23706, (2011).
[22] Y. Zhang, M. Belić, H. Zheng, H. Chen, C. Li, Y. Li, Y. Zhang, "Interactions of Airy beams, nonlinear accelerating beams, and induced solitons in Kerr and saturable nonlinear media" Opt. Exp **22**, 7160, (2014)
[23] R. Driben, Y. Hu, Z. Chen, B.A. Malomed, and R. Morandotti,, "Inversion and tight focusing of Airy pulses under the action of third-order dispersion," Opt. Lett. **38**, 2499 (2013).
[24] R. Driben, and T. Meier "Regeneration of Airy pulses in fiber-optic links with dispersion management of the two leading dispersion terms of opposite signs", Physical Review A **89**, 043817 (2014)
[25] Y. Zhang, M. Belić, Z. Wu, H. Zheng, K. Lu, Y. Li, Y. Zhang "Soliton pair generation in the interactions of Airy and nonlinear accelerating beams" Optics letters **38**, 4585 (2013)
[26] R. Driben, B. A. Malomed, A. V. Yulin, and D. V. Skryabin, "Newton's cradles in optics: From to N-soliton fission to soliton chains", Phys. Rev. A **87**, 063808 (2013)
[27] M. Mazilu, J. Baumgartl, T. Cizmar, and K. Dholakia, "Accelerating vortices in Airy beams", Proc. SPIE 7430, 74300C–1 (2009)
[28] H. T. Dai, Y. J. Liu, D. Luo, and X. W. Sun, "Propagation dynamics of an optical vortex imposed on an Airy beam," Opt. Lett. **35**, 4075 (2010).
[29] Y. Jiang, K. Huang, and X. Lu, "Propagation dynamics of abruptly autofocusing Airy beams with optical vortices," Opt. Exp.**20**, 18579 (2012).
[30]R. P. Chen, KH. Chew, and S.He. "Dynamic control of collapse in a vortex airy beam." Scientific Reports **3**, 1406, (2013).
[31] Y. Silberberg, "Collapse of optical pulses", Opt. Lett. **15**, 1282 (1990).
[32] Malomed, B. A., Mihalache, D., Wise, F. and L. Torner "Spatiotemporal solitons", J. Opt. B **7**, R53 (2005).
[33] D. Mihalache, "Linear and nonlinear light bullets: Recent theoretical and experimental studies", Rom. J. Phys. 57, 352 (2012).
[34] D. Mihalache, "Multidimensional localized structures in optics and Bose-Einstein condensates: A selection of recent studies", Rom. J. Phys. 59, 295 (2014).